# Understanding the multiple magnetic structures of the intermetallic compound NdMn$_{1.4}$Co$_{0.6}$Si$_2$

Madhumita Halder[1#], A. K. Bera[2†], Amit Kumar[2], L. Keller[3], and S. M. Yusuf[2*]

[1]*Department of Physics, Indian Institute of Technology Bombay, Mumbai 400076, India*

[2]*Solid State Physics Division, Bhabha Atomic Research Centre, Mumbai 400085, India*

[3]*Laboratory for Neutron Scattering, Paul Scherrer Institut, CH-5232 Villigen PSI, Switzerland*

## Abstract

Magnetic phases for the intermetallic compound NdMn$_{1.4}$Co$_{0.6}$Si$_2$ have been investigated at various temperatures by dc magnetization, neutron diffraction and neutron depolarization. Our study shows multiple magnetic phase transitions with temperature ($T$) over 1.5-300 K. In agreement with dc-magnetization and neutron depolarization results, the temperature dependence of the neutron diffraction patterns shows five distinct regions with different magnetic phases. These temperature regions are (i) $T \geq 215$ K, (ii) 215 K $> T \geq 50$ K, (iii) 50 K $> T \geq 40$ K, (iv) 40 K $> T > 15$ K, and (v) $T \leq 15$ K. The corresponding magnetic structures are paramagnetic, commensurate collinear antiferromagnetic (AFM-I), incommensurate AFM (AFM-II), mixed ferromagnetic and AFM (FM+AFM-II), and incommensurate AFM (AFM-II), respectively.

**Keywords:** Intermetallics, Magnetic structures, Neutron diffraction

Electronic mail: [#]mhalder@phy.iitb.ac.in, [*]smyusuf@barc.gov.in
[†] Present Address: Helmholtz Zentrum Berlin für Materialien und Energie, D-14109 Berlin, Germany

## 1. Introduction

The ternary rare earth $RT_2X_2$ intermetallic compounds ($R$: rare earth, $T$: transition metals, $X$: Si or Ge) are known to exhibit a wide variety of interesting physical properties, such as superconductivity, magnetism, heavy fermion, and Kondo effect [1-3]. There are even reports on the actinide based (U/Th)$T_2X_2$ intermetallic compounds [4-8]. These compounds have mostly the body-centered tetragonal ThCr$_2$Si$_2$-type crystal structure with space group $I4/mmm$, where $R$, $T$, and $X$ atoms occupy $2a$ (0, 0, 0), $4d$ (0, 1/2, 1/4), and $4e$ (0, 0, $z$) crystallographic positions, respectively. In this structure, $R$, $T$ and $X$ atoms are stacked in layers along the $c$-axis in $R$–$X$–$T$–$X$–$R$ sequence [9, 10]. The magnetic behavior of $RT_2X_2$ intermetallic compounds containing manganese is of great interest, because in these systems, among the two magnetic sublattices i.e. $R$ and $T$, the transition metal sublattice orders magnetically only for the compounds with $T$ = Mn [2, 3, 10]. The magnetic properties of $R$Mn$_2X_2$ compounds mainly depends on the intralayer (within the $ab$ plane) Mn-Mn distance, $d_{Mn-Mn}$. There is a critical value of $d_{Mn-Mn}$, (about 2.87 Å) below which the coupling between interlayer Mn moments is antiferromagnetic (AFM) and above this value, the coupling is ferromagnetic (FM) in nature [11-13].

NdMn$_2$Si$_2$ is known to be AFM below 380 K with a magnetic ordering of Mn sublattice [14]. Below 33 K, NdMn$_2$Si$_2$ is FM with ordered moments at both Nd and Mn sublattices [14]. The Mn sublattice is a canted FM at low temperature. The exchange interaction is mainly of the Ruderman-Kittel-Kasuya-Yosida (RKKY) type [14]. On the other hand, NdCo$_2$Si$_2$ is AFM with only Nd sublattice ordering and reported to undergo three AFM transitions [15] below 32 K, and it is also reported that the compound undergoes multiple metamagnetic transitions under an external magnetic field above 40 kOe at 4.2 K [15, 16].

Therefore, by replacing Mn in $NdMn_2Si_2$ by other $3d$ transition metals, e.g., Co (i.e. varying $T$), one can tune intraplane $T$-$T$ and interplane $R$-$T$ exchange interactions and achieve various interesting magnetic properties.

In our earlier report [17], we studied the magnetic properties and magnetocaloric effect in $NdMn_{2-x}Co_xSi_2$ series. Various interesting phenomena, such as metamagnetic transitions and domain wall pinning were observed for $NdMn_{2-x}Co_xSi_2$, and their role in obtaining a large magnetocaloric effect and an inverse magnetocaloric effect, respectively, was brought out [17]. Neutron diffraction study at 5 K for the $x = 0.2$ sample of this series, showed that the Nd moments were aligned along the crystallographic $c$-axis and the Mn moments were canted to the $c$-axis. The canted Mn moments were found to be coupled ferromagnetically along the $c$-axis and antiferromagnetically in the $ab$-plane. In this series, for the $x = 0.6$ sample i.e. $NdMn_{1.4}Co_{0.6}Si_2$, the magnetization data [17] showed multiple magnetic phase transitions with varying temperature. To understand the origin and nature of different magnetic phases, we have carried out a detailed dc-magnetization, neutron diffraction, and neutron depolarization study on the $NdMn_{1.4}Co_{0.6}Si_2$ compound, and report here the results.

## 2. Experimental Details

Polycrystalline sample of $NdMn_{1.4}Co_{0.6}Si_2$ was prepared by an arc-melting as described in our previous report [17]. The dc magnetization measurements were carried out on the sample using a Physical Property Measurement System (PPMS, Quantum Design) as a function of temperature and magnetic field. The zero-field-cooled (ZFC) and field-cooled (FC) magnetization measurements were carried out over the temperature range of 5–330 K in the presence of 300 Oe field. Neutron diffraction patterns were recorded at various temperatures over 1.5–300 K using the neutron powder diffractometer DMC ($\lambda$ = 2.4585 Å) at the Paul Scherrer Institute (PSI), Switzerland. The measured diffraction patterns were analyzed using the Rietveld refinement technique ( by employing the FULLPROF computer program [18]). The one-dimensional neutron-depolarization measurements were carried out down to 2 K using the polarized neutron spectrometer (PNS) at the Dhruva reactor ($\lambda$ = 1.205 Å), Trombay, Mumbai, India [19]. For FC neutron-depolarization measurements, the sample was first cooled from room temperature down to 2 K in the presence of 50 Oe field (required to maintain the neutron beam polarization at the sample position) and then the transmitted neutron beam polarization was measured as a function of sample temperature in warming cycle under the same field. The incident neutron beam was polarized along the z direction (vertically up) with a beam polarization of 98.60(1)% [19].

## 3. Results and discussion

Fig. 1 shows the ZFC and FC magnetization (*M*) vs temperature (*T*) curves under an applied field of 300 Oe for the compound $NdMn_{1.4}Co_{0.6}Si_2$ over the temperature range of 5 - 330 K. Multiple magnetic transitions are evident with the varying temperature. The FC magnetization decreases over 5 to 17 K, with increasing temperature and then increases over 17 to 24 K. It is seen that a clear magnetic phase transition occurs around 45 K. On a careful investigation, a broad peak is also observed around ~ 210 K, indicating a possible transition from AFM to paramagnetic state with increasing temperature. These multiple magnetic phase transitions are not present in pure $NdMn_2Si_2$ compound [14] or in compounds with small substitution of Co i.e in $NdMn_{2-x}Co_xSi_2$ with $x \leq 0.4$ [17].

In order to have a microscopic understanding of the multiple magnetic phase transitions, observed in the magnetization curve, we have carried out a detailed temperature dependent neutron diffraction study. Fig. 2 shows the neutron diffraction patterns for the compound at different temperatures over 1.5 – 300 K. The thermograms over $1.5 \leq T \leq 175$ K are characterized by the occurrence of several superlattice reflections and confirm the multiple magnetic transitions as a function of temperature. Diffraction patterns corresponding to the different magnetic phases are shown in different bunches. In order to investigate the magnetic phases further, the integrated intensity of a few selective nuclear as well as satellite peaks are plotted as a function of temperature in Fig. 3. The temperature dependence of the integrated peak intensities shows five distinct regions with changing temperature as; (i) $T \geq 215$ K, (ii) 215 K $> T \geq 50$ K, (iii) 50 K $> T \geq 40$ K, (iv) 40 K $> T > 15$ K, and (v) $T \leq 15$ K. We discuss below the nature of the phases across these five regions of temperature.

The neutron diffraction patterns at $T$ = 215 and 300 K could be fitted with only the nuclear phase confirming the paramagnetic state. These diffraction patterns (at 300 and 215 K) have been fitted with a tetragonal symmetry with a centrosymmetric space group $I4/mmm$ as reported for the compounds of the ThCr$_2$Si$_2$-type series *viz.* NdMn$_{2-x}$Co$_x$Si$_2$ [17]. The peak (at a scattering angle of $2\theta$ ~ 70 degree which is temperature independent [Fig. 4]) arises due to scattering from the sample holder, and therefore excluded from the analysis of the low temperature diffraction patterns. Within the tetragonal crystal structure with space group $I4/mmm$, the Mn/Co and Nd ions are located at the 4$d$ (0, 1/2, 1/4) and 2$a$ (0, 0, 0) sites, respectively. The Si atoms occupy the 4$e$ site (0, 0, $z$) with $z$ = 0. 3746(4) at 300 K. The lattice parameters are found to be $a = b = 3.9928(1)$ Å and $c = 10.3067(3)$ Å at 300 K.

Over the temperature range 215 K > $T \geq$ 50 K (175, 125, 100, 75, and 50 K), additional magnetic contribution appears at $2\theta$ = 53.7, 68.7, and 77.8 degrees. These additional magnetic peaks could be indexed with (111), (113) and (201) Bragg peaks, respectively, with respect to the tetragonal unit cell. The commensurate propagation vector $k_1$ = (001) has been used to index these magnetic satellite peaks. Under the body centered space group *I*4/*mmm*, the nuclear and magnetic reflections can obey the following conditions: (i) (*hkl*) with *h+k+l=2n* for the nuclear reflections, (ii) (*hkl*) with *h+k=2n* for the ferromagnetic ordering within the (00*l*) transition metal (Mn/Co) planes, (iii) (*hkl*) with *h+k=2n+1* for the antiferromagnetic ordering of the Mn moments within the (00*l*) Mn/Co planes, and (iv) (*hkl*) with *h+k+l=2n+1* for antiferromagnetic ordering between the adjacent Mn/Co-planes. For the present compound, all observed magnetic satellite peaks satisfy the condition *h+k* = 2*n* which are characteristics of the ferromagnetic ordering of transition metal (Mn) moments in *ab*-planes. The *C*-centered and anti *I*-center (111) line is characteristic of an antiferromagnetic coupling of such FM Mn layers yielding a net AFM ordering with collinear arrangement of Mn moments. The fitted neutron diffraction pattern at 75 K (a representative temperature in the range 215 K > $T \geq$ 50 K) and the corresponding magnetic structure is shown in Fig. 5. The magnetic moments are aligned along the crystallographic *c*-direction. The refined value of the Mn magnetic moment is found to be 1.98(6) $\mu_B$ at 75 K.

At 40 K, (in the temperature range 50 K > $T \geq$ 40 K), the magnetic satellite peaks at $2\theta$ = 53.7, 68.7, and 77.8 degree disappear and new satellite peaks appear at $2\theta$ = 37.1, 40.8, 52.5, 57.3, 60.7, 62.5, and 70.4 degrees. All peaks could be indexed with an incommensurate propagation vector $k_2$ = (0 0 0.345) with respect to the tetragonal crystal structure. The satellite peaks at $2\theta$ = (37.1, 40.8) and (52.5, 60.7) degrees are indexed as (101)±$k_2$ and (103)

±$k_2$, respectively. The peaks at 2$\theta$ = 57.3 and 62.5 degrees are indexed as (112)±$k_2$, and the peak at 2$\theta$ = 70.4 degree is indexed as (200)±$k_2$. The presence of the satellite peaks of the (10$l$) lines indicates the ($hkl$) rule with $h+k=2n+1$. The anti $C$-centered and $I$-centered lines (10$l$) with $l =2n+1$ (i.e., 1 and 3) are characteristics of an AFM ordering of the Mn spin within $ab$-plane. The propagation vector $k_2$ = (0 0 0.345) is along the $c$-axis. Therefore, the incommensurate component is along the $c$-axis. The observed satellite peaks to the (10$l$) lines, therefore, suggest a helical magnetic structure with an AFM arrangement of the Mn spins within the $ab$-plane where the axis of the helix is along the $c$-axis. The spins lie on the $ab$-plane. On the other hand, the (112) line is both $C$- and $I$-centered which is characteristic of the ferromagnetic (00$l$) Mn/Co planes (coupled ferromagnetically along the $c$-axis). Among (002) and (200) $C$- and $I$-centered peaks, the satellite lines are present for (200), not for (002) which suggests that the moments are aligned along the $c$-axis. Therefore, the net magnetic structure is an amplitude modulated AFM structure. The modulation is along the propagation vector direction i.e., along the $c$-axis. Since the moments are aligned along the $c$-axis, the amplitude modulated AFM structure also contributes intensity to the satellite peaks to another $C$-centered and $I$-centered line (110). However, no intensity to the satellite peaks (110) ± $k_2$ (2$\theta$ = 52 degree) has been observed which could be explained on the basis of a magnetic ordering of the Nd sub-lattice. The Nd ions are located at the special crystallographic position (000) in the unit cell. Owing to the absence of any intensity of the satellite peaks of the other $C$-centered and $I$-centered (002) line, it suggest that the Nd moments are also aligned along the $c$-axis (the propagation vector direction). Therefore, it suggests an amplitude modulated AFM arrangement for the Nd sublattices too. Owing to the positions of the Mn/Co and Nd ions, the magnetic structure factor of the line (110) is given by the difference between the magnetizations of the transition metal and Nd sublattices, which justify the absence of intensity to the (110) ± $k_2$ peaks. The observed neutron diffraction pattern along with the calculated diffraction pattern considering the magnetic ordering of both sublattices is shown in Fig. 6. The resultant magnetic structure for the transition metal sublattice is an amplitude modulated conical AFM structure, whereas for the Nd sublattice, it is an amplitude modulated collinear AFM structure (Fig. 6). The refined values of the magnetic moments are found to be (i) $\mu_{Mn-ab}$ = 1.17(8) $\mu_B$, (ii) $\mu_{Mn-c}$ = 2.0(1) $\mu_B$, and (iii) $\mu_{Nd}$ = 2.18(7) $\mu_B$. Here we would like to mention that in addition to these magnetic peaks another weak satellite peak also appears at $2\theta$ = 47.2 degree which can be indexed with (003)+$k_2$. However, the peak (003) is not allowed by the centrosymmetric space group where (00$l$) peaks with $l$ = 2$n$ ($n$ is an integer number) are only allowed. Perhaps the mixing of Mn/Co leads to a small preferred placement of the Mn/Co ions at different positions disturbing the $I$-centering. This disturbance of the symmetry is not strong enough to be visible in the nuclear data.

At 20 K, in the temperature range 40 K > $T$ > 15 K, in addition to the magnetic satellite peaks to the (10$l$), (103), (112), and (200) lines (as observed at 40 K), an increase in the intensity of the nuclear peaks (101) and (112) has been observed. Besides, a decrease in the intensity of the satellite peaks to the (001), (10$l$), (112) lines have also been observed. The (001) peak which is forbidden in the centrosymmetric space group, has also been observed, indicating the breaking of the body center symmetry. The increase in the intensity of the nuclear peaks (101) and (112) suggests a FM ordering of both transition metal and Nd sub-lattices. This is consistent with magnetization data (shown in Fig. 1) and neutron depolarization data (presented later) where a FM phase has been found over ~ 40 - 15 K temperature range. The refinement of the full diffraction pattern at 20 K (which includes all the above mentioned Bragg peaks) suggests that a net ferromagnetic ordering exists in addition to the AFM structure as found at 40 K. The ferromagnetic moment is found to align along the crystallographic *c*-axis. A reduction of the amplitude of the AFM component is evident from the decrease in the intensity of the satellite peaks. The refined neutron diffraction pattern and the corresponding magnetic structure are shown in Fig. 7. The refined values of the magnetic moments are found to be (i) $\mu_{Mn-ab}$ = 0.3(2) $\mu_B$, $\mu_{Mn-c}$ = 1.2(1) $\mu_B$ and (ii) $\mu_{Nd}$ = 2.58(8) $\mu_B$ for the AFM component, and (iii) $\mu_{Mn}$= 1.10(6) $\mu_B$, $\mu_{Nd}$ = 2.60(4) $\mu_B$ for the FM component.

In the temperature region $T \leq 15$ K (15, 10 and 1.5 K), the FM (101) and (112) peaks disappear. The forbidden (001) peak disappears as well. The intensity of the satellite peaks (001), (10l), (112) enhances. The AFM state, as found at 40 K, is reentered in this temperature range. The refined neutron diffraction pattern at 1.5 K, and the corresponding magnetic structure are shown in Fig. 8. At 1.5 K, the refined values of the magnetic moments are found to be (i) $\mu_{Mn-ab}$ = 1.52(8) $\mu_B$, (ii) $\mu_{Mn-c}$ = 2.25(9) $\mu_B$, and (iii) $\mu_{Nd}$ = 3.03(3) $\mu_B$, respectively.

The presence of the FM phase over the intermediate temperature region (40 K > $T$ > 15 K) has been supported by our one dimensional neutron depolarization study. Fig. 9 shows the temperature dependence of the transmitted neutron beam polarization *P* for the NdMn$_{1.4}$Co$_{0.6}$Si$_2$ compound. Four temperature regions with varying *P* are observed in the neutron depolarization data. At $T \geq 40$ K, no neutron depolarization has been observed. The value of *P* first decreases (depolarization) below 40 K, and attains its minimum value at ~ 26 K. With further lowering of the temperature, the polarization increases at $T < 26$ K and there in no depolarization below ~ 15 K. The absent of any depolarization above 40 K suggests a paramagnetic or an AFM state. Above 215 K, sample is paramagnetic, as confirmed from the refinement of the neutron diffraction data. Hence no depolarization is observed as in a paramagnetic material, the neutron polarization is unable to follow the variation in the internal magnetic field as the temporal spin fluctuation is too fast ($10^{-12}$ s or faster) [20-28]. In the temperature region 215 K > $T \geq$ 50 K, the sample is a collinear AFM as obtained from the neutron diffraction data (Fig. 5). In pure AFM materials, there is no net magnetization even at the magnetic domain level, hence no depolarization occurs [20-28]. No depolarization is found down to 40 K as the magnetic structure is amplitude modulated AFM at 40 K (with no net magnetization). Below 40 K, the value of $P$ first decreases down to 26 K which suggests an occurrence of net ferromagnetic moment. Further lowering of the temperature, the $P$ value increases and depolarization completely disappears only below ~ 15 K. This is consistent with neutron diffraction results where a net ferromagnetic moment appeared over 15 < $T$ < 40 K in addition to the incommensurate AFM structure. In the case of a ferromagnetic material, the magnetic domains exert a dipolar field on the neutron polarization and depolarize the neutrons due to Larmor precession of the neutron spins in the magnetic field of domains [20-28]. No depolarization is observed at $T \leq$ 15 K which confirms the presence of only an AFM phase below this temperature. This is in agreement with the neutron diffraction study which shows the re-stabilization of the incommensurate antiferromagnetic structure at $T \leq$ 15 K.

We discuss below the possible origin of the multiple magnetic phase transitions as a function of temperature, as found in the present study, for the $NdMn_{1.4}Co_{0.6}Si_2$ compound. The parent compound $NdMn_2Si_2$ is AFM below 380 K because of negative interlayer Mn-Mn coupling and canted FM below 33 K because of positive interlayer Nd - Nd and Nd - Mn couplings [14]. In this compound, the magnetic ordering of Nd layer results in a reorientation of Mn spins which leads to a canted FM structure of Mn spins below 33 K. On the other hand,

NdCo$_2$Si$_2$ is a collinear AFM below 15 K, and undergoes a series of magnetic phase transitions between 15 - 32 K with a square wave structure [15, 29]. Here, only Nd atoms order magnetically whereas Co atoms do not order. For NdCo$_2$Si$_2$ system, in the temperature region 24 K $< T <$ 32 K, the propagation vector is (0 0 0.785), whereas in the temperature region 15 K $< T <$ 24 K, the propagation vector is (0 0 0.928) [15, 29]. For temperature below 15 K, the propagation vector is (0 0 1) [15, 29]. When the Co concentration is less in NdMn$_{2-x}$Co$_x$Si$_2$ series, i.e., $x <$ 0.6, the system overall behaves like NdMn$_2$Si$_2$ (e.g., in case of $x =$ 0.2 and 0.4) [17]. Further replacing Mn by Co, i.e., $x =$ 0.6 leads to complex magnetic phases. Below 50 K, in case of the present sample NdMn$_{1.4}$Co$_{0.6}$Si$_2$, we have three distinct temperature regions, (as in case of NdCo$_2$Si$_2$) each having different magnetic structures. Similar multiple magnetic phase transitions with temperature has also been reported for the NdMn$_{2-x}$Fe$_x$Ge$_2$ series by Venturini *et al* [30]. They observed that the FM-AFM transition does not only depend on Mn-Mn distance, $d_{Mn-Mn}$, but the electronic effects also determine the magnetic behaviour of these compounds. For the present NdMn$_{1.4}$Co$_{0.6}$Si$_2$ compound, we find that the $d_{Mn-Mn}$ distance is less than the critical value (about 2.87 Å) throughout the investigated temperature range of 1.5–300 K (Fig. 10) for the interplane ferromagnetic structure, reported for pure Mn compounds. Therefore an AFM ordering of the Mn moments is expected throughout the temperature range below ~ 175 K. However, below 50 K, when Nd sublattice also orders, it leads to complicated magnetic structures. In case of NdMn$_2$Si$_2$, when the Nd sublattice orders ferromagnetically at low temperature, it reorients the Mn spin to a canted FM structure [14]. In case of NdCo$_2$Si$_2$, only Nd sublattice orders antiferromagnetically below 32 K with three distinct temperature regions 24 K $< T <$ 32 K, 15 K $< T <$ 24 K, and $T <$ 15 K. Thus when the transition metal is Mn, the Nd sublattice shows

FM ordering, whereas when the transition metal is Co, the Nd sublattice shows an AFM ordering. Thus, on substituting Mn by Co, there seems to be a competition between FM ordering and incommensurate AFM ordering of the Nd sublattice. This results in an amplitude modulated AFM structure. The Nd sublattice in $NdCo_2Si_2$ undergo three distinct AFM ordering with temperature [15], similar to the present $NdMn_{1.4}Co_{0.6}Si_2$ compound, where we find three distinct temperature regions below 50 K i.e the below ordering temperature of Nd atoms.

## 4. Summary and conclusion

In summary, neutron diffraction, neutron depolarization, and dc-magnetization techniques have been employed to investigate the multiple magnetic phases for the intermetallic compound $NdMn_{1.4}Co_{0.6}Si_2$ as a function of temperature. Five distinct magnetic phases are found: (i) In the temperature region above 215 K, the system is in the paramagnetic state. (ii) Over the temperature range 50 - 215 K, the system is in a commensurate collinear antiferromagnetic state (AFM-I). (iii) Over the temperature region 40 - 50 K, magnetic structure is a complex incommensurate AFM (AFM-II) comprising of an amplitude modulated conical AFM structure for the Mn sublattice and an amplitude modulated collinear AFM structure for Nd sublattice. (iv) In the temperature region 40 K > $T$ > 15 K, there is an existence of mixed AFM-II and FM states. (v) At $T \leq 15$ K, the FM phase disappears and the incommensurate AFM-II structure-stabilizes (as found over 40 - 50 K). The dc magnetization and neutron depolarization data support these temperature dependent multiple magnetic phase transitions. Nd site is found to order only below 50 K, and contributes distinctly to the observed three magnetic phases below 50 K.

**List of Figures**

Fig. 1. (Color online) Temperature dependence of FC and ZFC magnetization ($M$) for $NdMn_{1.4}Co_{0.6}Si_2$ sample plotted on logarithmic scale at 300 Oe applied field. The magnetic phase transitions are marked with arrow.

Fig. 2. (Color online) Neutron diffraction patterns for $NdMn_{1.4}Co_{0.6}Si_2$ compound at different temperatures. Different bunch of patterns are used to show the different magnetic phases.

Fig. 3. (Color online) Temperature dependence of the integrated intensity of the selective nuclear and satellite peaks. Five different regions of magnetic phases, as probed by neutron diffraction, are also shown.

Fig. 4. (Color online) Experimentally observed (circles) and calculated (solid line) neutron diffraction patterns for $NdMn_{1.4}Co_{0.6}Si_2$ compound at 300 K. The difference between observed and calculated patterns is shown by solid lines at the bottom. The vertical bars indicate the position of allowed nuclear Bragg peaks. The arrow mark indicates the peak which arises due to scattering from the sample holder.

Fig. 5. (Color online) Refined neutron diffraction pattern for $NdMn_{1.4}Co_{0.6}Si_2$ compound at 75 K and the corresponding magnetic structure. The arrow mark indicates the peak position which arises due to scattering from the sample holder.

Fig. 6. (Color online) Refined neutron diffraction pattern for $NdMn_{1.4}Co_{0.6}Si_2$ compound at 40 K. The corresponding magnetic structure (for one unit cell as well as and four unit cells) is also depicted. The arrow mark indicates the peak position which arises due to scattering from the sample holder.

Fig. 7. (Color online) Refined neutron diffraction pattern for NdMn$_{1.4}$Co$_{0.6}$Si$_2$ compound at 20 K and the corresponding magnetic structure with one as well as four unit cells. The arrow mark indicates the peak position which arises due to scattering from the sample holder.

Fig. 8. (Color online) Refined neutron diffraction pattern for NdMn$_{1.4}$Co$_{0.6}$Si$_2$ compound at 1.5 K and the corresponding magnetic structure with one as well as four unit cells. The arrow mark indicates the peak position which arises due to scattering from the sample holder.

Fig. 9. Temperature dependence of the transmitted neutron beam polarization $P$ at an applied field of 50 Oe for NdMn$_{1.4}$Co$_{0.6}$Si$_2$ compound.

Fig. 10. (Color online) Variation of lattice constants $a$ and $c$, and $d_{Mn-Mn}$ with temperature for NdMn$_{1.4}$Co$_{0.6}$Si$_2$ compound. The solid lines are guide to eye.

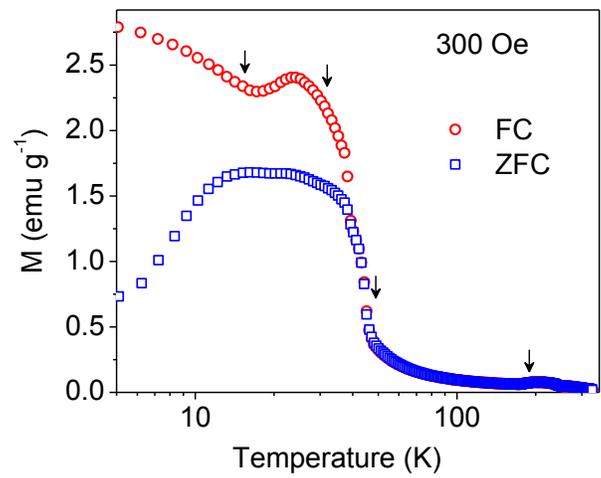

Fig. 1
Halder *et al*.

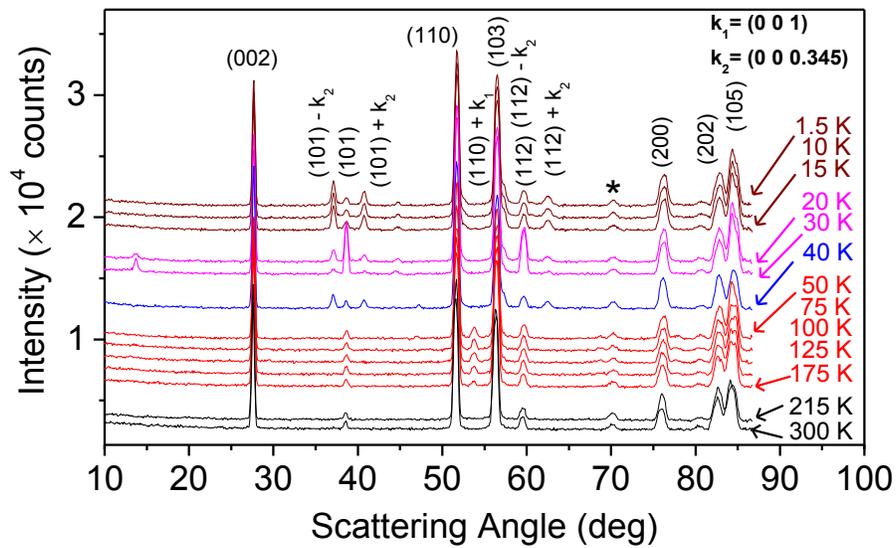

Fig. 2
Halder *et al*.

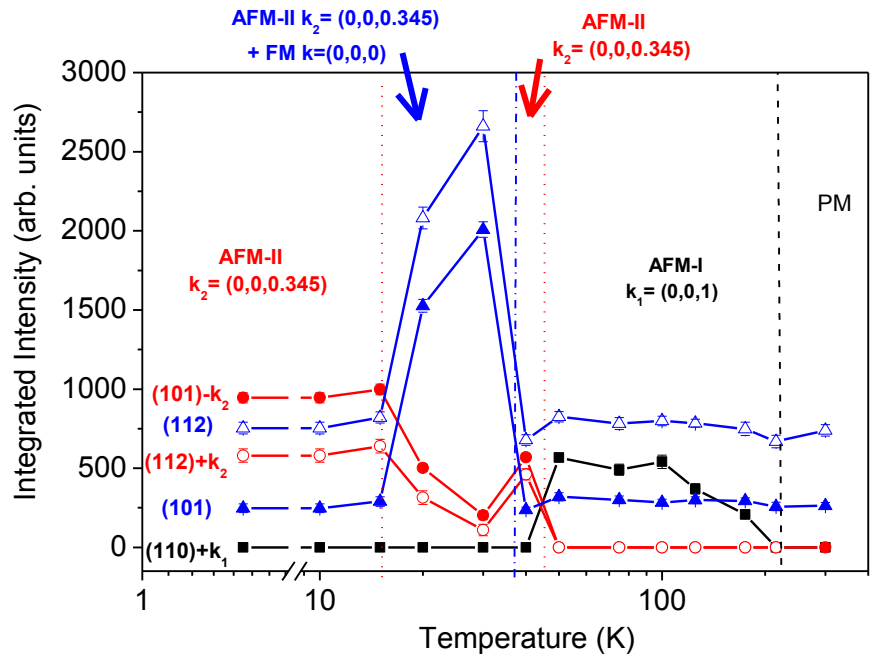

Fig. 3
Halder *et al*.

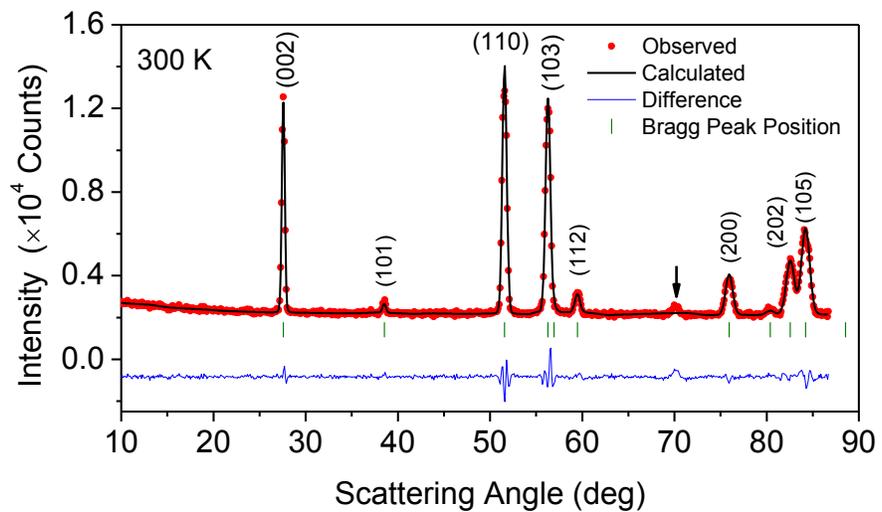

Fig. 4
Halder *et al*.

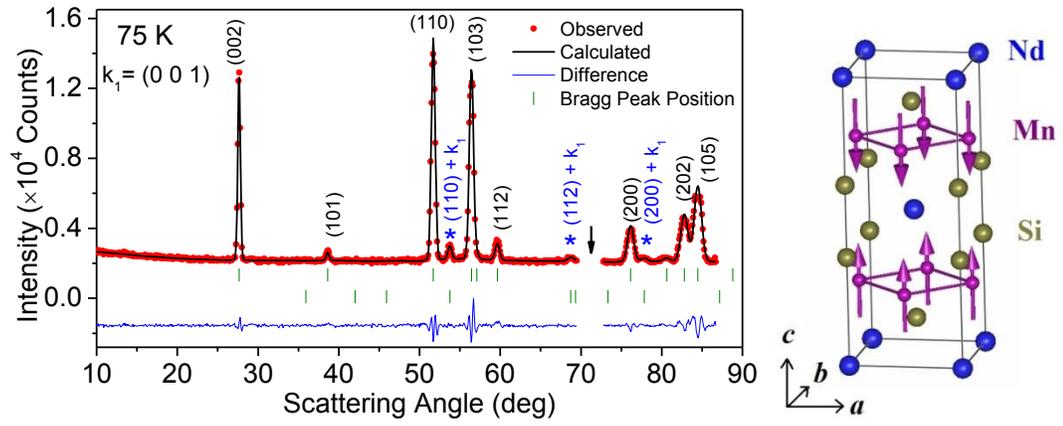

Fig. 5
Halder *et al*.

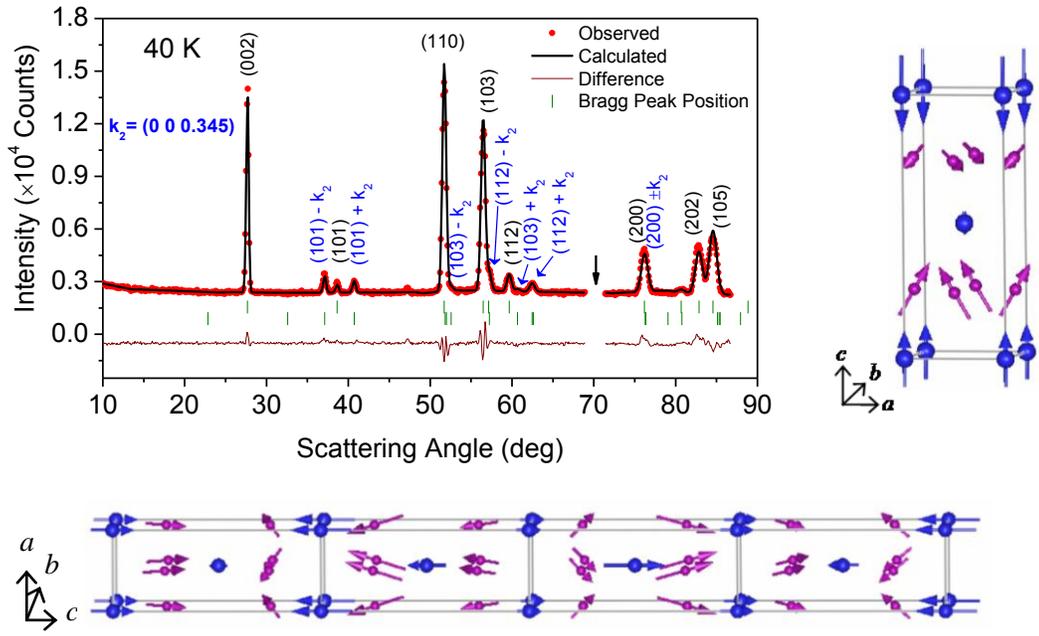

Fig. 6
Halder *et al*.

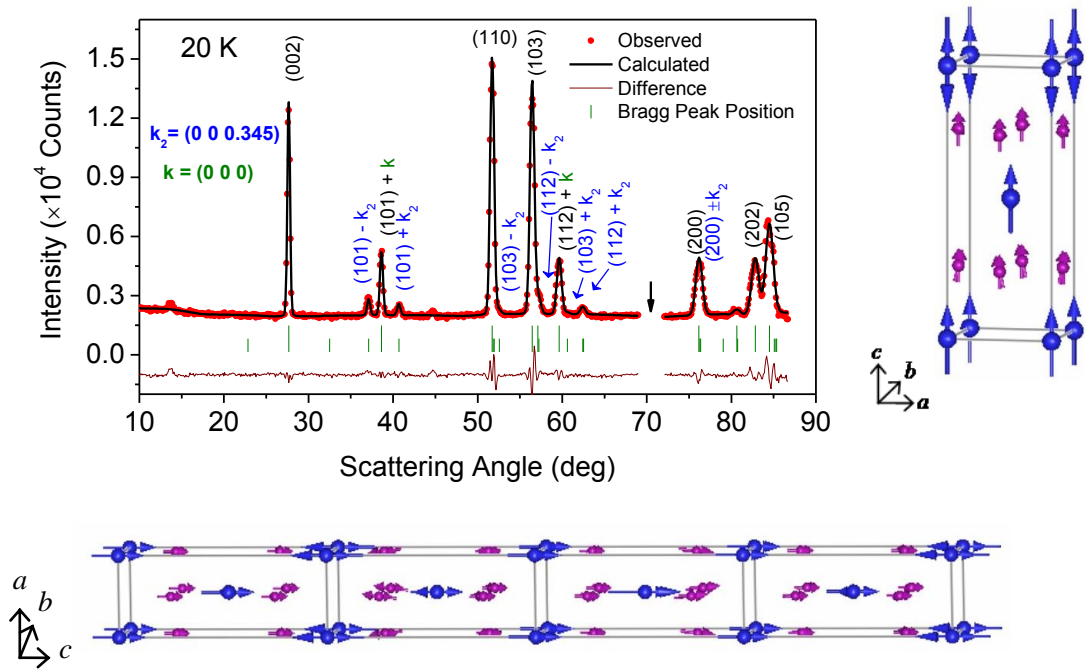

Fig. 7
Halder *et al*.

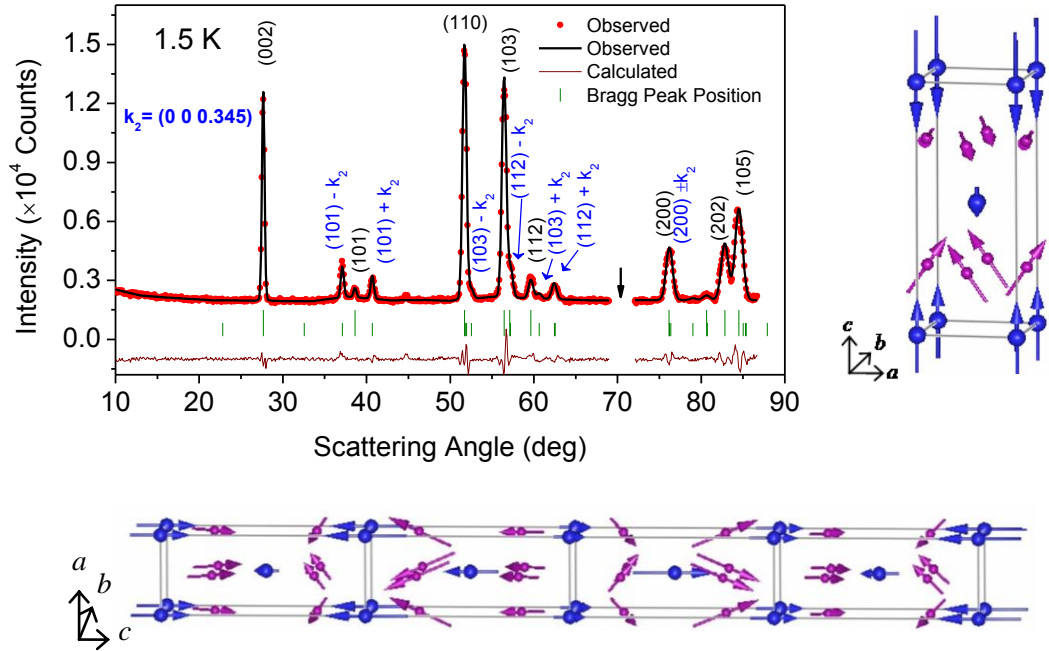

Fig. 8
Halder *et al*.

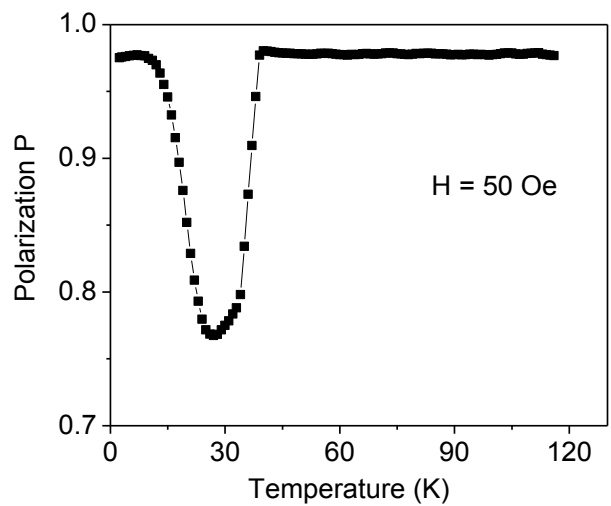

Fig. 9
Halder *et al*.

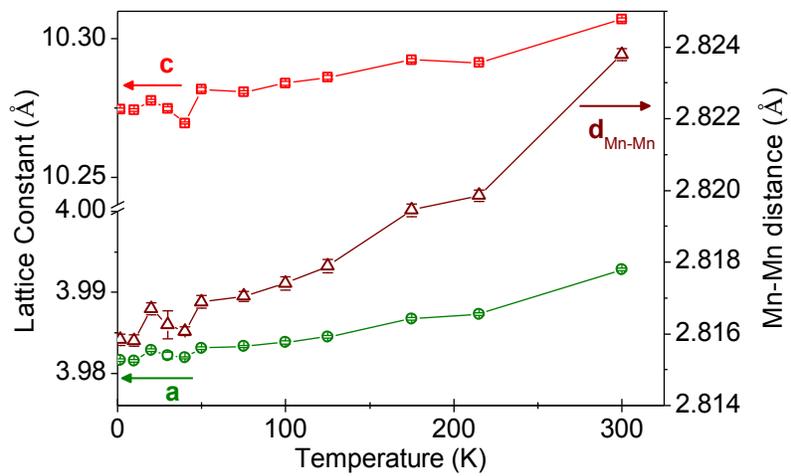

Fig. 10
Halder *et al*.